\documentclass[sigplan,screen, authorversion, nonacm]{acmart}
\AtBeginDocument{%
  \providecommand\BibTeX{{%
    \normalfont B\kern-0.5em{\scshape i\kern-0.25em b}\kern-0.8em\TeX}}}

\setcopyright{acmcopyright}
\copyrightyear{2018}
\acmYear{2018}
\acmDOI{10.1145/1122445.1122456}

\acmConference[Woodstock '18]{Woodstock '18: ACM Symposium on Neural
  Gaze Detection}{June 03--05, 2018}{Woodstock, NY}
\acmBooktitle{Woodstock '18: ACM Symposium on Neural Gaze Detection,
  June 03--05, 2018, Woodstock, NY}
\acmPrice{15.00}
\acmISBN{978-1-4503-XXXX-X/18/06}

\usepackage[algo2e,ruled,vlined,noend]{algorithm2e}
\SetCommentSty{textit}
\DontPrintSemicolon
\IncMargin{-\parindent}
\SetAlCapHSkip{0pt}
\SetAlgoLined

\SetKwFor{ParallelFor}{for}{do in parallel}{}
\SetKwIF{If}{ElseIf}{Else}{if}{}{else if}{else}{end if}

\usepackage{pifont}

\newcommand{\Partition}{\ensuremath{\mathrm{\Pi}}}%
\newcommand{\incnets}{\ensuremath{\mathrm{I}}}%
\newcommand{\pinsinpart}{\ensuremath{\mathrm{\Phi}}}

\newcommand{\netweight}{\ensuremath{\mathrm{\omega}}}
\newcommand{\vertexweight}{\ensuremath{\mathrm{c}}}

\newcommand{\con}{\ensuremath{\lambda}}
\newcommand{\conset}{\ensuremath{\Lambda}}

\newcommand{\pluseq}{\mathrel{+}=}
\newcommand{\minuseq}{\mathrel{-}=}
\newcommand{\argmax}{\operatorname{arg \, max}}
\newcommand{\gain}{\operatorname{gain}}

\newcommand{\vol}{\operatorname{vol}}
\newcommand{\cov}{\operatorname{cov}}

\newcommand{\splitatcommas}[1]{%
  \begingroup
  \begingroup\lccode`~=`, \lowercase{\endgroup
    \edef~{\mathchar\the\mathcode`, \penalty0 \noexpand\hspace{0pt plus 1em}}%
  }\mathcode`,="8000 #1%
  \endgroup
}

\renewcommand{\epsilon}{\varepsilon}

\begin{document}

\title{Deterministic Parallel Hypergraph Partitioning}

\author{Lars Gottesbüren}
\email{lars.gottesbueren@kit.edu}
\orcid{1234-5678-9012}
\affiliation{%
  \institution{Karlsruhe Institute of Technology}
  \streetaddress{Am Fasanengarten 5}
  \city{Karlsruhe}
  \country{Germany}
  \postcode{76135}
}
\author{Michael Hamann}
\email{michael.hamann@kit.edu}
\orcid{0000-0002-6958-4927}
\affiliation{%
	\institution{Karlsruhe Institute of Technology}
	\streetaddress{Am Fasanengarten 5}
	\city{Karlsruhe}
	\country{Germany}
	\postcode{76135}
}

\begin{abstract}
Balanced hypergraph partitioning is a classical NP-hard optimization problem with applications in various domains such as VLSI design, simulating quantum circuits, optimizing data placement in distributed databases or minimizing communication volume in high-performance computing.
Engineering parallel heuristics for this problem is a topic of recent research.
Most of them are non-deterministic though.
In this work, we design and implement a highly scalable deterministic algorithm in the state-of-the-art parallel partitioning framework Mt-KaHyPar.
On our extensive set of benchmark instances, it achieves similar partition quality and performance as a comparable but non-deterministic configuration of Mt-KaHyPar and outperforms the only other existing parallel deterministic algorithm BiPart regarding partition quality, running time and parallel speedups.

\end{abstract}



\maketitle

\section{Introduction}

Balanced k-way hypergraph partitioning (HGP) is a classical optimization problem.
Its goal is to divide the vertices of a hypergraph into $k$ blocks of bounded size while minimizing an objective function on hyperedges that connect more than one block.
Hypergraphs are a generalization of graphs where hyperedges can contain more than two vertices.
A commonly used objective function is the \emph{connectivity metric} where we aim to minimize the sum of the number of blocks connected by each hyperedge.
This problem is NP-hard~\cite{LENGAUER} and even hard to approximate~\cite{AndreevRaeckeApprox} within constant factors, which is why heuristic algorithms are used in practice.
Balanced hypergraph partitioning has numerous applications in domains such as VLSI design~\cite{ISPD98, ALPERT-SURVEY}, logic synthesis for integrated circuits~\cite{LogicSynthesisApplication}, simulating quantum circuits~\cite{gray2021hyper}, scientific computing~\cite{PATOH}, SAT solving~\cite{MANN-PAPA14}, as well as storage sharding in distributed databases and data centers~\cite{SHP, schism}.

There has been a huge amount of research on graph and hypergraph partitioning, but especially in recent years, the interest in parallel algorithms has surged~\cite{mt-kahypar-d, mt-kahypar-q, deep-mgp, BIPART, SHP, ZOLTAN} due to ever growing problem sizes.
With the exception of \texttt{BiPart}~\cite{BIPART}, these algorithms are non-deterministic.
Researchers have advocated the benefits of deterministic parallel algorithms for several decades~\cite{DBLP:conf/popl/Steele90, DBLP:journals/computer/Lee06, bocchino2009parallel, DBLP:conf/ppopp/BlellochFGS12}, including ease of debugging, reasoning about performance, and reproduciblity.
While some strive for deterministic programming models, we want to leverage randomized scheduling for better performance and thus pursue deterministic algorithms.
The above reasons are desirable properties, yet the \texttt{BiPart} authors~\cite{BIPART} argue that for VLSI circuit design, deterministic partitioning results are even \emph{necessary}, since some manual post-processing is involved that should not be repeated.

Previous results showed that the quality of partitions computed by \texttt{BiPart} does not compare favorably with current state-of-the-art parallel algorithms such as \texttt{Mt-KaHyPar}~\cite{mt-kahypar-q} and our experiments show that \texttt{BiPart} exhibits poor scalability.
The goal of this work is to design, implement and evaluate a scalable and deterministically parallel hypergraph partitioning algorithm with state-of-the-art solution quality.

\subsection{Multilevel Partitioning}

More formally, the hypergraph partitioning problem is defined as follows.
Given a hypergraph $H=(V,E)$, \emph{imbalance parameter} $\epsilon \in (0,1)$, and number of blocks $k \in \mathbb{N}$, find a $k$-way partition $V_1 \cup \dots \cup V_k = V$ of the vertices $V$ that is $\epsilon$-balanced $|V_i| \leq (1 + \epsilon) \lceil |V|/k \rceil$.
The objective function to minimize is the \emph{connectivity metric} $\sum_{e \in E} (\lambda(e)-1)$, where $\lambda(e) := |\{V_i \mid V_i \cap e \neq \emptyset\}|$ is the number of blocks connected by hyperedge $e$.
The most successful approach for partitioning is the multilevel framework.
Starting with the input hypergraph, a sequence of successively smaller (coarser) but structurally similar hypergraphs is constructed by repeatedly contracting groups of vertices (clusters or matchings).
This is called the \emph{coarsening phase}.
Once a sufficiently small hypergraph is reached, expensive algorithms compute an \emph{initial partition}.
In the following \emph{refinement phase}, this partition is then projected back up through the hierarchy by assigning vertices to the same block as their representative in the next coarser hypergraph.
This yields a partition of the larger hypergraph with the same imbalance and objective function as on the coarse hypergraph.
On each level of the hierarchy, local search algorithms such as Fiduccia-Mattheyses (FM)~\cite{FM} or label propagation~\cite{LABEL_PROPAGATION} move vertices to improve the current partition.

\texttt{Mt-KaHyPar}~\cite{mt-kahypar-d} adds a preprocessing phase based on community detection to the coarsening phase.
Community detection, also called clustering, aims to group vertices together that are internally densely but externally sparsely connected.
Contractions during the coarsening phase are then restricted to vertices in the same community to avoid destroying small cuts, as first proposed in~\cite{KAHYPAR-CA}.
To avoid confusion, we use the terms \emph{community detection} and \emph{communities} in the preprocessing phase, while we use \emph{clustering} and \emph{clusters} in the coarsening phase.

There are two fundamental approaches to the multilevel framework.
In direct $k$-way partitioning, the initial partition is directly $k$-way and refinement algorithms operate on a $k$-way partition.
Recursive bipartitioning instead obtains a $k$-way partition by recursively splitting the blocks in two parts.
This is simpler as it only requires refinement algorithms that work on two-way partitions.
Most partitioners~\cite{PATOH, ZOLTAN, HMETIS, BIPART}, including \texttt{BiPart} use recursive bipartitioning.
However, recursive bipartitioning often achieves substantially worse solution quality than direct $k$-way~\cite{SimonTeng97}, which is why we focus on direct $k$-way in this paper.

\subsection{Non-Determinism in Local Moving}

\begin{algorithm2e}[t]
	\SetEndCharOfAlgoLine{}
	\caption{Local Moving Round}\label{algo:async-local-moving}
	\ParallelFor() {$v \in V$ in random order} {
		compute and perform best move for $v$ \;
		update data structures \;
	}
\end{algorithm2e}

Typical community detection, clustering coarsening and label propagation refinement algorithms are so-called \emph{local moving algorithms}, which follow the structure outlined in Algorithm~\ref{algo:async-local-moving}: given an initial assignment of vertices to groups, visit vertices in random order in parallel, and improve the solution by greedily moving vertices when they are visited.
Since vertices are moved right away, the local optimization decisions depend on non-deterministic scheduling decisions.
Our approach to incorporate determinism is illustrated in Algorithm~\ref{algo:sync-local-moving}.
It is based on the \emph{synchronous local moving} approach of Hamann et al.~\cite{SLM} to parallelize the Louvain community detection algorithm~\cite{Louvain} on distributed memory.

Instead of performing moves asynchronously, vertices are split into sub-rounds -- using deterministically reproducible randomness.
In each sub-round, the best move for each vertex in the current sub-round is computed with respect to the unmodified groups.
In a second step, some of the calculated moves are approved and performed, and some are denied, for example due to the balance constraint.

\begin{algorithm2e}[t]
	\SetEndCharOfAlgoLine{}
	\caption{Synchronous Local Moving Round}\label{algo:sync-local-moving}
	randomly split vertices into sub-rounds \;
	\For() {$r=0$ \KwTo number of sub-rounds} {
		\ParallelFor() {$v \in V$ in sub-round $r$} {
			compute and save best move for $v$\;
		}
		approve saved moves and update data structures
	}
\end{algorithm2e}

\subsection{Contributions}

We propose three deterministic parallel local moving algorithms: for the preprocessing, coarsening and refinement phases of \texttt{Mt-KaHyPar}.
An algorithmic novelty is the incorporation of vertex weights in the approval step of the refinement phase via a merge-style parallelization.
Extensive experiments demonstrate that our new algorithm achieves good speedups (28.7 geometric mean, 48.9 max on 64 threads) and achieves similar solution quality and performance as its non-deterministic counterpart (as expected slightly worse overall though).
We investigate potential causes for this cost of determinism, finding that the coarsening phase is the most affected.
Our algorithm outperforms \texttt{BiPart} regarding solution quality, running time and parallel speedups on 98\% of the instances.

Our shared-memory implementation is available as open-source software as part of the \texttt{Mt-KaHyPar} framework.
We describe engineering details, and also describe how to incorporate determinism in the rest of the framework.

The paper is organized as follows.
In Section~\ref{sec:preliminaries}, we introduce concepts and notation.
Subsequently, we describe our algorithmic components and the implementation in Section~\ref{sec:main}.
In Section~\ref{sec:experiments}, we analyze the algorithm experimentally via a parameter study and comparison with existing algorithms, before concluding the paper in Section~\ref{sec:conclusion}.

\section{Preliminaries}\label{sec:preliminaries}

By $[m]$ we denote the set $\{0, 1, \dots, m-1\}$ for a positive integer $m$.
We use Python-style slicing notation $A[i:j]$ to denote sub-arrays from index $i$ up to (excluding) index $j$.

\paragraph{Hypergraphs}

A \emph{weighted hypergraph} $H=(V,E,c,\omega)$ is a set of vertices $V$ and a set of hyperedges $E$ with vertex weights $c:V \to \mathbb{N}$ and hyperedge weights $\omega:E \to \mathbb{N}$, where each hyperedge $e$ is a subset of the vertex set $V$.
The vertices of a hyperedge are called its \emph{pins}.
A vertex $v$ is \emph{incident} to a hyperedge $e$ if $v \in e$.
$\incnets(v)$ denotes the set of all incident hyperedges of $v$.
The \emph{degree} of a vertex $v$ is $d(v) := |\incnets(v)|$.
The \emph{size} $|e|$ of a hyperedge $e$ is the number of its pins.
By $N(v) := \{ u \in V \mid \incnets(u) \cap \incnets(v) \neq \emptyset\}$ we denote the \emph{neighbors} of $v$.
We extend $c$ and $\omega$ to sets in the natural way $c(U) :=\sum_{v\in U} c(v)$ and $\omega(F) :=\sum_{e \in F} \omega(e)$.

A graph is a hypergraph where each edge has size $2$.
We use the terms nodes and edges when referring to graphs, and vertices, hyperedges and pins when referring to hypergraphs.

\paragraph{Partitions}

A \emph{$k$-way partition} of a hypergraph $H$ is a function $\Partition : V \to [k]$ that assigns each vertex to a \emph{block} (or block identifier) $i \in [k]$.
In addition to the identifiers, we also use the term block for their corresponding vertex sets $V_i := \Partition^{-1}(i)$.
We call $\Partition$ \emph{$\varepsilon$-balanced} if each block $V_i$ satisfies the \emph{balance constraint}:
$c(V_i) \leq L_{\max} := (1+\varepsilon) \lceil c(V) / k \rceil$ for some parameter $\varepsilon \in (0,1)$.

For each hyperedge $e$, $\conset(e) := \{V_i \mid  V_i \cap e \neq \emptyset\}$ denotes the \emph{connectivity set} of $e$.
The \emph{connectivity} $\con(e)$ of a hyperedge $e$ is $\con(e) := |\conset(e)|$.
A hyperedge is called a \emph{cut hyperedge} if $\con(e) > 1$.
Given parameters $\varepsilon$, and $k$, and a hypergraph $H$, the \emph{hypergraph partitioning problem} is to find an $\varepsilon$-balanced $k$-way partition $\Partition$ that minimizes the \emph{connectivity metric} $(\lambda - 1)(\Pi) := \sum_{e \in E} (\lambda(e) - 1) \: \omega(e)$.
We use the term solution quality for connectivity metric in the experiments.
In order to avoid confusion later on, note that, while these concepts are defined as functions, the pseudocodes in this paper treat them as data that are modified as the algorithms iteratively change the partition.

\paragraph{Contraction}

A \emph{clustering} (or set of \emph{communities}) $\mathcal{C}$ is a partition without a restriction on the number of clusters (vertex blocks).
\emph{Contracting} a clustering yields a smaller, \emph{coarser} hypergraph, by merging vertices assigned to the same cluster into a super-vertex whose weight is the sum of the constituents.
Correspondingly, the hyperedges of the coarse hypergraph become smaller and duplicate hyperedges may arise.
Duplicates are removed and their weight is aggregated at the sole non-removed representative.

Given a partition $\Partition_c$ of a hypergraph $H_c$ that was obtained from another hypergraph $H$ by contracting a clustering $\mathcal{C}$, the partition $\Partition_c$ can be projected to $H$ by assigning vertices to the block of their super-vertex $\Partition(v) = \Partition_c(\mathcal{C}(v))$.
The partition of $H$ then has the same imbalance and objective function value as that of $H_c$.
This is the fundamental property of the multilevel framework.

\paragraph{Parallel Primitives}

We use parallel primitives such as prefix sums, reduce and sorting from Intel's \texttt{tbb} library.
Whenever possible, we use our own counting sort (integer sorting) implementation~\cite[Section 8.2]{clrs-ia-01} with $O(\log(n) + K)$ depth and $O(n + K)$ work, where $n$ is the size of the input, and $K$ is the maximum key value.

A fundamental primitive in our algorithms is deterministic parallel random shuffling, or more often, randomly splitting elements into sub-rounds.
Our algorithm is based on the RandSort and RandDist algorithms~\cite{DBLP:conf/ISCApdcs/CongB05}.
We divide the input range into a fixed number of equal-size chunks (256 regardless of number of threads used for reproducibility), which are handled independently in parallel.
For each chunk, a random number generator is seeded with an input seed and the index of the first element, which is then used to generate 8-bit tags for each element in the chunk.
The input range is then sorted by the tags using parallel counting sort.
For splitting into sub-rounds, the algorithm stops here -- counting sort returns an array of offsets where each tag starts.
Multiple tags are further grouped together to form sub-rounds as needed.
For random shuffling, the elements with the same tag are shuffled sequentially, but each tag independently in parallel.
The order within the same tag is deterministic because counting sort is stable.

\section{Deterministic Parallel Multilevel Partitioning}\label{sec:main}

In this section, we describe our algorithms.
The structure follows the order of the multilevel framework.
We introduce the community detection preprocessing in Section~\ref{sec:preprocessing}, the coarsening in Section~\ref{sec:coarsening}, the initial partitioning in Section~\ref{sec:initial-partitioning}, and the direct $k$-way refinement in Section~\ref{sec:refinement}.

\subsection{Preprocessing}\label{sec:preprocessing}

The preprocessing phase detects communities in the hypergraph that are used to guide the coarsening process by restricting contractions to vertices in the same community, as proposed in~\cite{KAHYPAR-CA}.
We perform community detection on the bipartite graph representation of the input hypergraph with some modified edge weights to handle large hyperedges.
The bipartite graph representation $G=(V_G, E_G)$ of a hypergraph $H=(V,E)$ has the vertices and hyperedges as nodes, and an edge for every pin connecting the vertex and hyperedge.
Since this also assigns hyperedges to communities, we subsequently restrict the communities to the vertices.
The edge weights are set to $w'(v,e) := \frac{\omega(e) |\incnets(v)|}{|e|}$ or $w'(v,e) := \omega(e)$ for pin $v \in e$ and hyperedge $e \in E$, depending on the density of the hypergraph as described in~\cite{KAHYPAR-CA}.

We heuristically maximize the well-known modularity objective using a synchronous parallel version of the popular Louvain algorithm~\cite{Louvain}.
The modularity of given communities $\mathcal{C}$ is $\mathcal{Q(C)} := \cov(\mathcal{C}) - \sum_{C \in \mathcal{C}} {\vol(C)}^2 / {\vol(V_G)}^2$.
Here, the coverage $\cov(C) := \sum_{C \in \mathcal{C}} \sum_{u\in C}\sum_{v \in C \cap N(u)} w'(u,v) / \vol(V_G)$ is the fraction of edge weights inside communities.
The volume $\vol(u) := \sum_{v \in N(u)} w'(u,v)$ is the sum of incident edge weights of a node (counting self-loops twice), which is extended to node-sets $\vol(X) := \sum_{u \in X} \vol(u)$.

The Louvain algorithm starts with each node in its own community.
In a round, it visits each node in a random order and greedily maximizes modularity by possibly moving the node to the community of a neighbor.
After a fixed number of rounds or if no node has been moved in the last round, the communities are contracted and the algorithm is applied recursively to the contracted graph.
This continues until no node has been moved on a level, at which point the community assignment is projected to the input graph.

The gain (modularity difference) of moving a node from its current community to a neighboring community can be computed purely from the weight of incident edges to the target or current community, as well as their volumes.
Therefore, it suffices to store and update the volume for each community, and compute the weights to the communities by iterating once over the neighbors of the node.

We randomize the visit order by dividing nodes into random sub-rounds.
For each sub-round we calculate the best move for each node in parallel, but only apply volume and community assignment updates after synchronizing, to make the algorithm deterministic.

\paragraph{Volume Updates.}

One intricacy with updating the community volumes is that adding floating point numbers is not commutative, and thus the previous approach~\cite{PARALLEL-LOUVAIN, mt-kahypar-d} of applying all updates in parallel with compare-and-swap instructions is non-deterministic.
Instead, we have to establish an order in which the volume updates of each community are aggregated.
For this, we collect all necessary updates in a global vector, which we lexicographically sort by community (primary key) and node ID (secondary key).
Applying the updates is done in parallel for different communities.
To reduce the sorting overhead we split the updates into two vectors (addition and subtraction) which are sorted independently in parallel, but applied one after another.

To analyze the work and depth, let $V'_G$ denote the nodes in a sub-round.
The work is $\sum_{u \in V'_G} \deg(u) + |V'_G|\log(|V'_G|)$.
The depth is linear in the maximum number of moves in or out of a community (sequential volume updates) and the maximum degree $\max_{u \in V'_G}(\deg(u))$ for calculating modularity gains, plus the depth of the sorting algorithm.
This is usually poly-logarithmic in $|V'_G|$, but \texttt{tbb}'s quick-sort implementation uses sequential partitioning, and thus is linear.
We tested a supposedly better sorting algorithm but did not achieve an improvement.
The number of moves and degree linear terms in the depth may be reduced to poly-logarithmic by parallelizing the per-vertex gain calculation (parallel for loop over neighbors, atomic fetch-add for weight to neighbor clusters) and aggregating updates within a community in parallel with a deterministic reduce.
However, in practice the outer level of parallelism is sufficient.

\paragraph{Contraction.}

To contract the communities, we first remap the community IDs to a consecutive range.
This is done by computing the prefix sum over an array with $1$'s in the locations of used community IDs.
In a second pass, we remap the node-to-community assignment by looking up the old community ID in the prefix sum array.

Subsequently, we sort the nodes by community ID via counting sort, which also gives us an array of offsets into the sorted range where each community begins.
We parallelize the generation of the coarse edges on a per-community level.
For each community, one thread generates all of its outgoing coarse edges by iterating sequentially over the neighbors of the vertices in the community, and aggregating the edge weights in a vector indexed by community ID of the neighbor.
Since the sorting algorithm is stable, the orders in which node volumes are aggregated and edges are generated are deterministic.
The work is $O(|E_G|)$, and the depth is the largest degree sum over vertices in a community.

\subsection{Coarsening}\label{sec:coarsening}

After performing community detection on the bipartite graph representation, we proceed to contracting the actual hypergraph.
In the coarsening phase we keep performing coarsening passes over the vertices (Algorithm~\ref{algo:coarsening-pass}) until only a few vertices remain, at which point we can run initial partitioning.
We choose this contraction limit $CL = 160 \cdot k$ dependent on $k$, as in~\cite{KAHYPAR-K, mt-kahypar-d}.
In each coarsening pass, we perform one round of local moving and then contract the resulting clusters.
The objective function for the clustering is the commonly used~\cite{PATOH, KAHYPAR-K, mt-kahypar-d} heavy-edge rating function $r(u, C) := \sum_{e \in I(u) \cap I(C)} \frac{\omega{(e)}}{|e| - 1}$ which rewards heavy hyperedges between a vertex $u$ and a potential target cluster $C$, but penalizes large hyperedges.

\begin{algorithm2e}[t]
	\KwIn{vertex $u \in V$}
	\SetEndCharOfAlgoLine{}
	\caption{Compute Heavy-Edge Rating}\label{algo:heavy-edge-rating}
	candidates $\gets \emptyset$ \;
	\For() {$e \in \incnets(u)$} {
		\For() {$v \in e$} {
			\If() {community$[u] = $ community$[v]$} {
				\If() {rating$[\mathcal{C}[v]] = 0$} {
					add $\mathcal{C}[v]$ to candidates
				}
				rating$[\mathcal{C}[v]] \pluseq \omega(e) / |e|$
			}
		}
	}
	\For() {$C \in$ candidates} {
		\If() {weight$[C] + c(u) \leq CW_{\max}$ and rating$[C] > $ best rating} {
			store $C$ as best candidate \;
		}
		rating$[C] \gets 0$ \;
	}
	\Return best candidate
\end{algorithm2e}

Initially, the clustering $\mathcal{C}$ is a singleton clustering, i.e., each vertex is in its own cluster.
For each vertex $u$ in a sub-round, we store the best target cluster according to the rating function in an array of propositions $\mathcal{P}$.
Algorithm~\ref{algo:heavy-edge-rating} shows pseudocode for calculating the ratings and Algorithm~\ref{algo:coarsening-pass} shows pseudocode for one coarsening pass over the vertices.
First, we aggregate the ratings in a sparse array indexed by cluster ID and store the potential candidates in a dense vector, before we select the highest-rated candidate and reset the ratings.
To save running time, and since their contribution to the rating function is small, we skip hyperedges with size $> 1000$.
This ensures that at most $O(|\incnets(v)|)$ time is spent for vertex $v$ (though the constant is large) instead of $O(\sum_{e \in \incnets(v)} |e|)$, which leads to work linear in the number of pins per coarsening pass to compute target clusters.
If there are multiple candidates with the same rating, we pick one uniformly at random -- to achieve deterministic selection we use a hash-and-combine function seeded with $u$ as a random number generator.
In Algorithm~\ref{algo:heavy-edge-rating}, we could theoretically parallelize the iteration over neighbors by using atomic fetch-and-add instructions for aggregating the ratings.
The check \textit{rating}$[\mathcal{C}[v]] = 0$ can be faithfully implemented because the atomic instruction returns the value immediately prior to its execution.
However, in practice the outer level of parallelism over the vertices is again sufficient.

\paragraph{Approving Moves.}

Since the initial partitioning step must be able to compute a feasible partition, we enforce a maximum weight on the clusters $CW_{\max} := \min \left( L_{\max}, c(V)/CL \right)$.
To respect this constraint, we filter the target cluster candidates further during the selection.
Additionally, some of the calculated moves must be rejected.
Therefore, we sort the moves lexicographically by cluster, vertex weight, and lastly vertex ID (for determinism).
For each target cluster, we then approve the vertices one by one (in order of ascending weight), and reject all of the remaining moves into this cluster once $CW_{\max}$ would be exceeded.
Our implementation iterates over the moves in parallel, and the iteration of the first vertex in the sub-range of a cluster is responsible for performing the moves into the cluster.
To drastically reduce the number of moves we have to sort, we employ an optimization, where we already sum up the cluster weights during the target-cluster calculation step using atomic fetch-and-add instructions, and simply approve all moves into a target cluster whose weight will not exceed $CW_{\max}$.
Due to this optimization, calculating the target clusters is by far the more expensive step in practice, even though approving the moves requires sorting.

\begin{algorithm2e}[t]
	\SetEndCharOfAlgoLine{}
	\caption{Coarsening Pass}\label{algo:coarsening-pass}
	randomly split vertices into sub-rounds \;
	$\mathcal{C}[u] \gets u, \mathcal{P}[u] \gets u \, : \, \forall u \in V$ \;
	opportunistic-weight$[u] \gets c(u) \, : \, \forall u \in V$ \;
	\For() {$r=0$ \KwTo number of sub-rounds} {
		\ParallelFor() {$u \in V$ in sub-round $r$} {
			$\mathcal{P}[u] \gets $ \FuncSty{ComputeHeavyEdgeRating(u)} \;
			opportunistic-weight$[\mathcal{P}[u]] \pluseq c(u)$ \tcp*[f]{atomic}
		}
		$M \gets \emptyset$ \tcp*[r]{moves}
		\ParallelFor() {$u \in V$ in sub-round $r$} {
			\If() { opportunistic-weight$[\mathcal{P}[u]] \leq CW_{\max}$ } {
				$\mathcal{C}[u] \gets \mathcal{P}[u]$ \;
			} \Else () {
				add $u$ to $M$ \;
			}
		}
		sort $M$ lexicographically by $(\mathcal{P}[u], c(u), u)$\;
		\ParallelFor() { $i=0$ \KwTo $|M|$ } {
			\If() {$i=0$ or\, $\mathcal{P}[M[i-1]] \neq \mathcal{P}[M[i]]$} {
				\For() {$j= i$ until $CW_{\max}$ exceeded} {
					$\mathcal{C}[M[j]] \gets \mathcal{P}[M[j]]$
				}
				set opportunistic-weight of $\mathcal{C}[M[i-1]]$
			}
		}
	}
	contract clustering $\mathcal{C}$ \;
\end{algorithm2e}

\paragraph{Contraction.}

The hypergraph contraction algorithm consists of several steps: remapping cluster IDs to a consecutive range (as for graph contraction), generating pin lists of the hyperedges of the contracted hypergraph, removing duplicate hyperedges, and finally assembling the data structure.

We generate the coarse pin list of each hyperedge in parallel, by replacing the vertex ID with the remapped cluster ID and removing duplicate entries.
Our version uses a bit-set for de-duplication, but sorting is not much slower.
At this stage we already discard hyperedges of size one.

To remove duplicate hyperedges, we use a parallel version of the INRSort algorithm~\cite{INR, INR-Source}.
The INRSort algorithm works as follows.
Comparing all hyperedge-pairs for equality is too expensive, so a hash function is used to restrict comparisons to hyperedges with equal hash value and size.
For parallelism, hyperedges are distributed across threads using the hash value of their pins~\cite{mt-kahypar-d}.
Each thread sorts its hyperedges by their hash value, their size, as well as ID for determinism.
In each sub-range with equal hash value and size (consecutive in memory due to sorting), pair-wise comparisons of their pins are performed.
Again, we use a bit-set to check for equality, as this was slightly faster than sorting the pins.
The running time of the de-duplication algorithm is difficult to analyze, since it depends on the collision rate of the hash function, the number of duplicate hyperedges and their sizes.
However, in practice, it is faster than constructing the pin lists and the incident hyperedge lists.

At this point, we have obtained the pin lists of the coarse hypergraph, and now need to construct the list of incident hyperedges at each vertex.
For this, we first count the number of incident hyperedges at each vertex, and compute a prefix sum over these values.
In a second pass, we write the incident hyperedges into the sub-ranges of the corresponding pins, using an atomic fetch-and-add instruction on the starting position of the sub-range.
Finally, we sort the incident hyperedges of each vertex for determinism.

\subsection{Initial Partitioning}\label{sec:initial-partitioning}

After the coarsening phase, we compute an initial $k$-way partition on the coarsest hypergraph.
We perform recursive bipartitioning with the multilevel algorithm and thus only need to provide \emph{flat} algorithms for computing initial $2$-way partitions.
Since the coarsest hypergraphs are \emph{small}, a portfolio of 9 different simple, sequential algorithms~\cite{KaHyPar-R} is used.
Combined with 20 repetitions each for diversity, there is ample parallelism.
Each run is followed up with 3 rounds of sequential FM local search~\cite{FM}.
These algorithms are inherently deterministic, however care must be taken when selecting which partition to use for refinement.
The primary criteria are connectivity followed by imbalance.
To achieve deterministic selection, we assign sequentially generated tags to the initial bipartitions, which are used as a tie breaking mechanism.
In combination with deterministic coarsening and refinement, the overall initial partitioning phase is deterministic.


\subsection{Refinement}\label{sec:refinement}

In the refinement phase, we take an existing $k$-way partition (from the previous level or initial partitioning) and try to improve it by moving vertices to different parts, depending on their gain values.
The gain of moving vertex $u \in V$ from its current block $s$ to block $t$ is
$\gain(u,t):= \sum_{e \in \incnets(u) : \pinsinpart(e, s) = 1}\omega(e) - \sum_{e \in \incnets(u) : \pinsinpart(e, t) = 0} \omega(e)$.
The first term accounts for the hyperedges $e$ for which $s$ will be removed from their connectivity set $\conset(e)$, the second term accounts for those where $t$ will be newly added.

\paragraph{Finding Moves.}

\begin{algorithm2e}[t]
	\KwIn{vertex $v \in V$}
	\SetEndCharOfAlgoLine{}
	\caption{Compute Max Gain Move}\label{algo:compute-km1-gain}
	gains$[i] \gets 0 \, \forall i \in [k]$ \;
	internal $\gets 0$ \;
	\For() {$e \in \incnets(v)$} {
		\If() {$\pinsinpart(e,\Partition[v]) > 1$} {
			internal $\pluseq \netweight(e)$
		}
		\For() {block $i \in \conset(e)$} {
			gains$[i] \pluseq \netweight(e)$
		}
	}
	$j \gets \argmax_{i \in [k]}(\text{gains}[i])$ \;
	\Return $j, \text{gains}[j] - \text{internal}$
\end{algorithm2e}

Our refinement algorithm is a synchronous version of label propagation refinement~\cite{PARHIP, mt-kahypar-d}.
The vertices are randomly split into sub-rounds.
For each vertex in the current sub-round, we compute the highest gain move, and store it if the gain is positive.
Algorithm~\ref{algo:compute-km1-gain} shows pseudocode for computing the gains of a vertex $v$ to all $k$ blocks, and selecting the highest gain move.
As an optimization it uses the connectivity sets $\conset(e)$ instead of checking the pin counts $\pinsinpart(e, i)$ for each block $i \in [k]$ directly.
The gain-calculation phase takes $O(k|V| + \sum_{u \in V} \sum_{e \in \incnets(u)} \lambda(e))$ work (across all sub-rounds) and $O(k + \max_{u \in V}(\sum_{e \in \incnets(u)} \lambda(e)))$ depth for each sub-round.
The $O(k)$ term per vertex for initializing the gains array and selecting the highest gain can be eliminated by tracking occupied slots and resetting only these, though this is not useful in practice if $k$ is small.

In a second step we approve some of the stored moves, and subsequently apply them in parallel, before proceeding to the next sub-round.
This is the interesting part, as just applying all moves does not guarantee a balanced partition.

\paragraph{Maintaining Balance By Vertex Swaps.}

In this step we perform a sequence of balance-preserving vertex swaps on each block-pair, prioritized by gain.
This approach was first introduced in \texttt{SHP}~\cite{SHP}, though their work only considers unweighted vertices as \texttt{SHP} is not a multilevel algorithm.
For each block-pair $(s,t) \in {[k]\choose 2}$, we collect the vertices $M_{st}$ that want to move from $s$ to $t$ and $M_{ts}$ from $t$ to $s$, and sort both sequences descendingly by gain (with vertex ID as tie breaker for determinism).
\texttt{SHP} now moves the first $\min(|M_{st}|, |M_{ts}|)$ vertices from each sequence.
If each vertex has unit weight, this does not change the balance of the partition.
However, we have to handle weights, so we are interested in the longest prefixes of $M_{st}, M_{ts}$, represented by indices $i,j$, whose cumulative vertex weights $c(M_{st}[0:i]), c(M_{ts}[0:j])$ are equal.
This is similar to merging two sorted arrays, as we describe in the next paragraph.
We do not have to swap exactly equal weight, as long as the resulting partition is still balanced.
For each block, we have a certain additional weight $B_t$ it can take before becoming overloaded.
If block-pairs are handled sequentially one after another, we can set $B_t = L_{\max} - c(V_t)$.
If they are handled in parallel, we divide this budget (equally) among the different block-pairs that have moves into $t$.

Again, we denote the prefixes as indices $i$ and $j$ into $M_{st}$ and $M_{ts}$, respectively.
The prefixes $i,j$ are called \emph{feasible} if they satisfy the condition $-B_s \leq c(M_{st}[0:i]) - c(M_{ts}[0:j]) \leq B_t$, i.e., swapping the first $i,j$ moves yields a balanced partition.
To compute the two longest feasible prefixes of $M_{st}, M_{ts}$, we can iterate simultaneously through both sequences and keep track of the so far exchanged vertex weight $c(M_{st}[0:i]) - c(M_{ts}[0:j])$.
If $c(M_{st}[0:i]) - c(M_{ts}[0:j]) < 0$ and $M_{st}$ has moves left, we approve the next move from $M_{st}$ by incrementing $i$.
Otherwise we approve the next move from $M_{ts}$.
In each iteration we check whether the current prefixes are feasible.

We parallelize this similar to a parallel merge algorithm.
In a first step, we compute cumulative vertex weights via parallel prefix sums.
Then the following algorithm is applied recursively.
We search for the cumulative weight of the middle of the longer sequence in the shorter sequence using binary search.
The left and right parts of the sequences can be searched independently.
If the right parts contain feasible prefixes, we return them, otherwise we return the result from the left parts.
The top-level recursive call on the left parts is guaranteed to find at least $i=j=0$ (no move applied).
If $n$ denotes the length of the longer sequence, this algorithm has $O(\log(n)^2)$ depth and $O(n)$ work.

Since we are interested in the longest prefixes, we can omit the recursive call on the left parts if the prefixes at the splitting points are feasible.
Depending on the available budgets $B_s, B_t$ this is fairly likely, since the cumulative weights are as close as possible.
Further, we can omit the recursive call on the right parts if the cumulative weight at the middle of the longer sequence exceeds the cumulative weight at the end of the shorter sequence plus the appropriate budget.
Note that in this case the binary search finds the end and thus the left recursion takes the entirety of the shorter sequence.

We stop the recursion and run the sequential algorithm if both sequences have less than $2000$ elements.
This value worked well in preliminary experiments.
As we already computed cumulative weights, we instead perform the simultaneous traversal from the ends of the sequences.
Since we expect to approve the majority of the saved moves, this is likely faster.

\paragraph{Further Implementation Details}

If we move vertices in parallel, the computed gains are not correct, as moving a vertex impacts the gains of its neighbors.
However, we can still obtain the sum of the exact gains using a technique dubbed attributed gains~\cite{mt-kahypar-d} which is outlined in Algorithm~\ref{algo:perform-move}, where we also describe the data structure updates incurred by a vertex move.
For each incident hyperedge $e$ of the moved vertex, we increment the pin count $\pinsinpart(e, t)$ in the target block $t$ and decrement $\pinsinpart(e, s)$ in the source block $s$.
If $\pinsinpart(e,s)$ becomes zero, we attribute an $\omega(e)$ connectivity improvement, similarly if $\pinsinpart(e, t)$ becomes one we attribute an $\omega(e)$ loss.
The sum of these attributed gains is equal to the overall improvement, though each individual attributed gain may not be correct.
If the overall attributed gain is negative, we made the solution worse.
Since we cannot single out particular bad moves, we instead revert to the partition before the sub-round.
A good strategy to avoid this in future iterations is to increase the number of sub-rounds for the current level.
However, this happens predominantly in the last round on the level, which is why we observed little impact.

\begin{algorithm2e}[t]
	\KwIn{vertex $v \in V$ to be moved from block $s$ to $t$}
	\SetEndCharOfAlgoLine{}
	\caption{Perform Move}\label{algo:perform-move}
	attributed $\gets 0$ \;
	\For() {$e \in \incnets(v)$} {
		lock$(e)$\;
		\If() {$(\pinsinpart(e, s) \minuseq 1) = 0$} {
			attributed $\pluseq \netweight(e)$ \;
			remove $s$ from $\conset(e)$ \;
		}
		\If() {$(\pinsinpart(e, t) \pluseq 1) = 1$} {
			attributed $\minuseq \netweight(e)$ \;
			add $t$ to $\conset(e)$ \;
		}
		unlock$(e)$ \;
	}
	$\Partition[v] \gets t$ \;
	part-weight$[s] \minuseq \vertexweight(v)$ \tcp*[r]{atomic}
	part-weight$[t] \pluseq \vertexweight(v)$ \tcp*[r]{atomic}
	\Return attributed
\end{algorithm2e}

To speed up the gain-calculation phase, we employ an optimization that is commonly known as \emph{active sets}.
We perform multiple rounds (consisting of sub-rounds) of the refinement algorithm.
Starting from the second round, only neighbors of vertices that were moved in the previous round are considered.
We implement this by activating neighbors of moved vertices, using compare-and-swap instructions on an array storing the last round in which a vertex was activated to avoid duplicate insertion.
To achieve linear work in the number of pins, we also use this mechanism to scan each hyperedge at most once per round.
At the beginning of the next round, the collected neighbors are sorted for determinism, before being split into sub-rounds.

\subsection{Differences to BiPart}

We now discuss differences between our algorithm and \texttt{BiPart}.
As already mentioned, \texttt{BiPart} uses recursive bipartitioning, whereas we use direct $k$-way, which is superior regarding solution quality~\cite{SimonTeng97}.
Other than $2$-way versus $k$-way and using sub-rounds and active sets, the refinement algorithms are largely the same, inspired by label propagation~\cite{PARHIP} and \texttt{SHP}~\cite{SHP}.
However, their refinement ignores vertex weights (apply all moves of the shorter sequence and the same number from the longer), which leads to imbalanced partitions that must be repaired by explicit rebalancing.
This can be slow and offers little control by how much solution quality degrades.
Our refinement guarantees balanced partitions at all stages without rebalancing.
Furthermore, \texttt{BiPart} uses no mechanism to track actual improvements, whereas we use attributed gains to detect and prevent quality-degrading moves.
Their coarsening scheme assigns each vertex to its smallest incident hyperedge (ties broken by ID) and merges all vertices assigned to the same hyperedge.
This offers no control over vertex weights and does not rank higher hyperedge weights as more important to not cut.
Preventing large vertex weights is important so that initial partitioning can find balanced partitions and there is more leeway for optimization in initial partitioning and refinement.
\texttt{BiPart} uses a parallel version of greedy graph growing~\cite{PATOH} for initial partitioning, even though the coarsest hypergraphs are small, where it is feasible to afford many diversified, parallel repetitions of sequential algorithms.

\section{Experiments}\label{sec:experiments}

\begin{figure*}
	\begin{minipage}{.33\textwidth}
		\includegraphics[width=\textwidth]{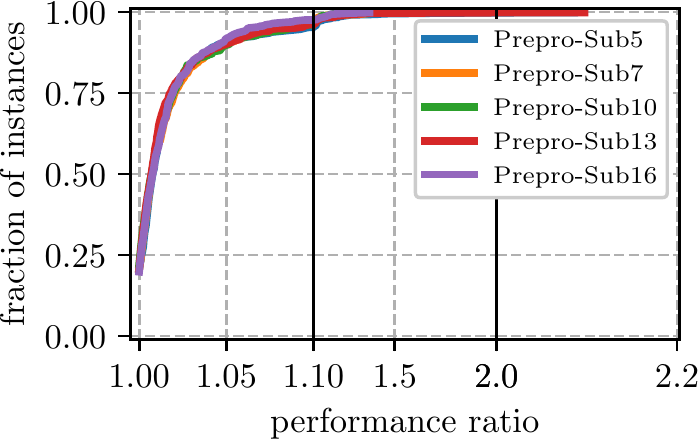}
	\end{minipage} %
	\begin{minipage}{.33\textwidth}
		\includegraphics[width=\textwidth]{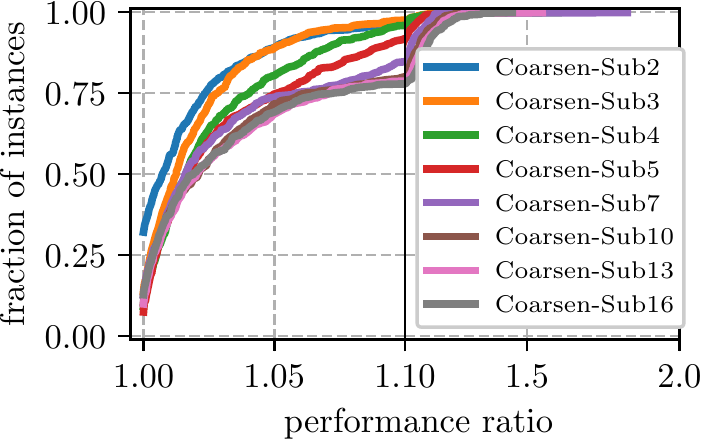}
	\end{minipage} %
	\begin{minipage}{.33\textwidth}
		\includegraphics[width=\textwidth]{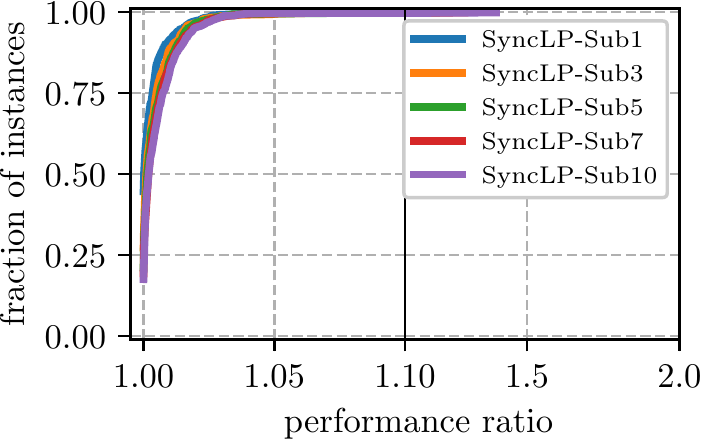}
	\end{minipage} %
	\caption{Performance profiles comparing the impact of the number of sub-rounds parameter on the different phases.}
	\label{fig:subrounds:quality}
\end{figure*}

\begin{figure}
	\includegraphics[width=0.66\linewidth]{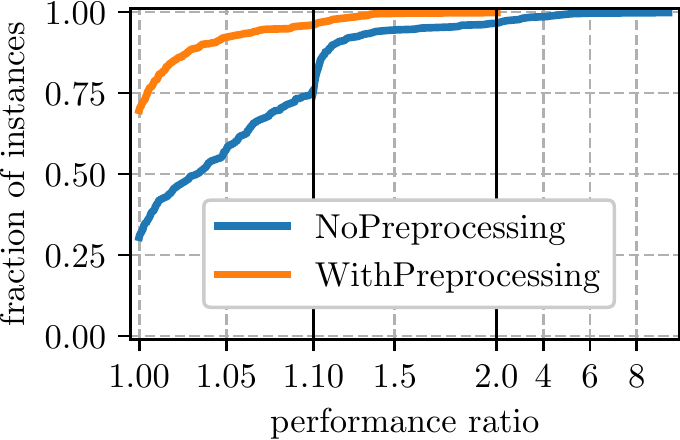}
	\caption{Solution quality with and without preprocessing.}\label{fig:quality:no-preprocessing}
\end{figure}

Our code is integrated in the \texttt{Mt-KaHyPar} hypergraph partitioning framework.
It is written in \texttt{C++17}, uses Intel's \texttt{tbb} library for parallelization and is compiled with \texttt{g++} version 9.2 with optimization level \texttt{-O3} and native architecture optimizations.
The experiments are run on a 128-core (2 sockets, 64 cores each) AMD EPYC Zen 2 7742 CPU clocked at 2.25GHz (3.4GHz turbo boost) with 1TB DDR4 RAM, and 256MB L3 cache.

\subsection{Benchmark Set}

We use the established benchmark set of 94 large hypergraphs that was assembled to evaluate \texttt{Mt-KaHyPar}, set B in~\cite{mt-kahypar-d}.
It contains hypergraphs from three different sources: VLSI instances from the DAC 2012 Routability-Driven Placement Contest~\cite{DAC}, various large sparse matrices from the SuiteSparse Matrix Collection~\cite{SPM}, and three different representations (literal, primal, dual) of SAT formulas from the 2014 SAT Competition~\cite{SAT14}.
We use $k \in \{2,4,8,16,32,64\}$, $\varepsilon = 0.03$, and partition each instance five times with different random seeds.
Since we have 94 instances, we cannot report instance sizes for each of them, however these statistics are available online\footnote{\url{https://algo2.iti.kit.edu/heuer/alenex21/}} in the supplementary material of~\cite{mt-kahypar-d}.
The largest instances have between $10^7$ and $10^8$ vertices and hyperedges, as well as $10^8$ to $2 \cdot 10^9$ pins.
Dual SAT instances are known for large hyperedges (with up to millions of pins), and hence their corresponding primal counterparts are known for large vertex degrees.
Some sparse matrices are even more skewed with a maximum degree of $10^7$.

\subsection{Configurations}

We perform $5$ rounds of local moving on each level during refinement, $5$ rounds before contracting during preprocessing, and one round before contracting during coarsening.
We call the algorithm and configuration proposed in this work \texttt{Mt-KaHyPar-SDet}, and the equivalent configuration that uses the existing non-deterministic local moving algorithms \texttt{Mt-KaHyPar-S}, where \texttt{Det} stands for determinism, and \texttt{S} for speed.
Additionally, we consider the configuration \texttt{Mt-KaHyPar-D} (for default) from~\cite{mt-kahypar-d}, which has a different initial partitioning configuration that is non-deterministic~\cite{mt-kahypar-q} and additionally uses parallel localized FM -- a more advanced refinement algorithm that is difficult to make deterministic.
The parallel n-level version from~\cite{mt-kahypar-q} is excluded here since it represents a vastly different time-quality trade-off.

\begin{figure*}
	\includegraphics[width=0.949\linewidth]{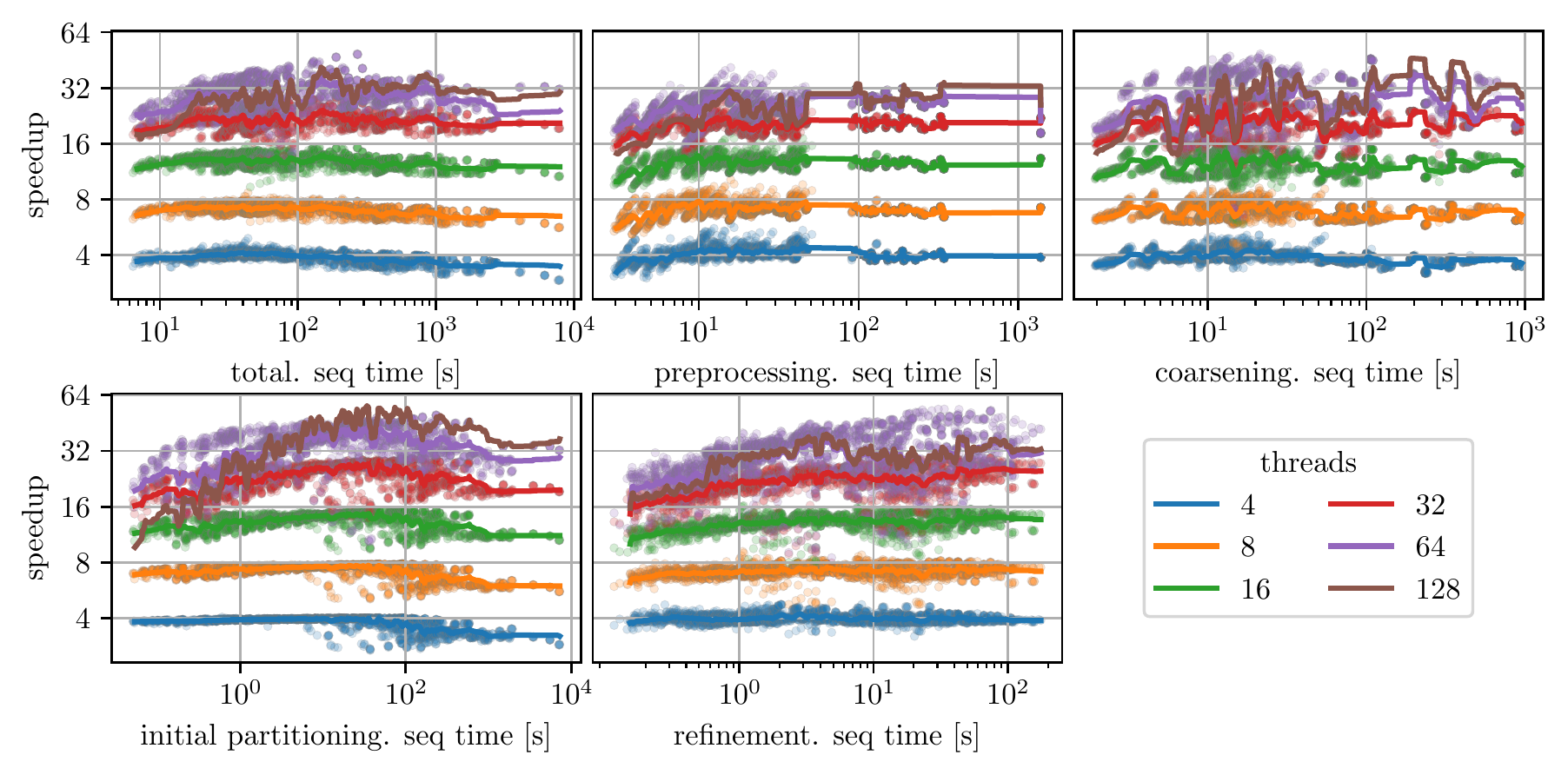}
	\caption{Speedups for \texttt{Mt-KaHyPar-SDet} in total as well as its components separately. The x-axis shows the sequential time in seconds, the y-axis the speedup. The lines are rolling geometric means (window size 50) of the per-instance speedups (scatter).}\label{fig:speedups:mt-kahypar-sdet}
\end{figure*}

\subsection{Performance Profiles}

To compare the solution quality of different algorithms, we use \emph{performance profiles}~\cite{PERFORMANCE-PROFILES}.
Let $\mathcal{A}$ be the set of all algorithms we want to compare, $\mathcal{I}$ the set of instances, and $q_{A}(I)$ the quality of algorithm
$A \in \mathcal{A}$ on instance $I \in \mathcal{I}$.
For each algorithm $A$, we plot the fraction of instances ($y$-axis) for which $q_A(I) \leq \tau \cdot \min_{A' \in \mathcal{A}}q_{A'}(I)$, where $\tau$ is on the $x$-axis.
Achieving higher fractions at equal $\tau$-values is considered better.
For $\tau = 1$, the $y$-value indicates the percentage of instances for which an algorithm performs best.
To interpret these plots, we either look at how quickly the curve converges towards $y=1$ (higher is better), or we look at the maximum ratio for certain instance fraction quantiles.
To calculate the ratios, we take the average connectivity across different seeds for each instance.
In addition to the plots, we report the $\tau$ ratios (performance ratios) at certain quantiles, as well as their geometric mean.

\subsection{Parameter Tuning}

\begin{figure}
	\includegraphics[width=\linewidth]{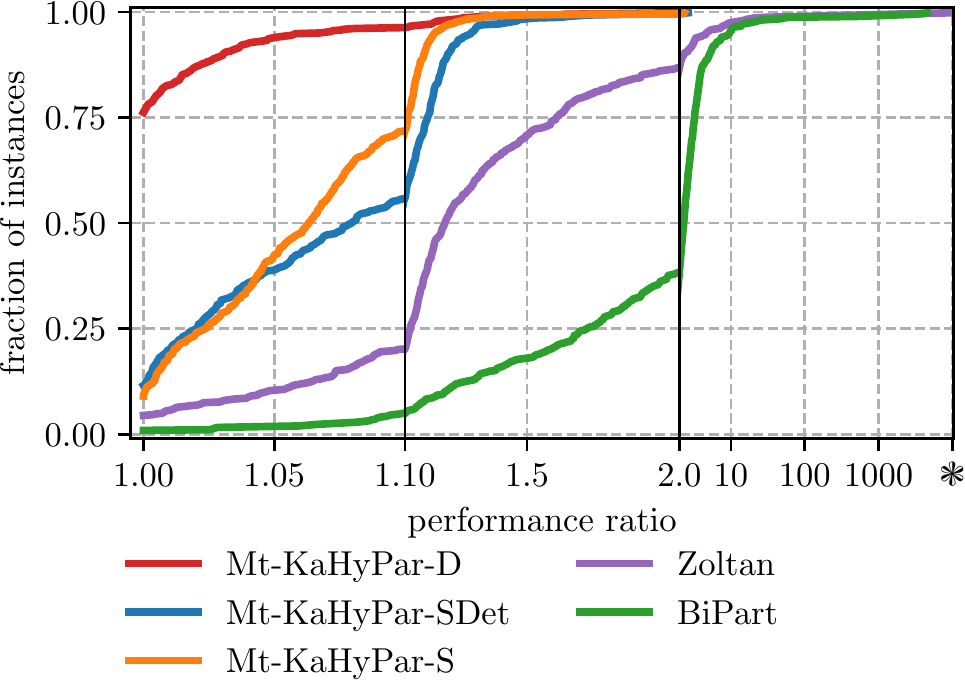}
	\caption{Performance profiles comparing the solution quality of \texttt{BiPart}, \texttt{Zoltan}, our algorithm \texttt{Mt-KaHyPar-SDet} and the existing \texttt{Mt-KaHyPar} variants. The \ding{99} symbol marks segmentation faults (6 instances for \texttt{Zoltan}).}\label{fig:quality:all-algos}
\end{figure}

\begin{figure*}
	\begin{minipage}{.33\textwidth}
		\includegraphics[width=\textwidth]{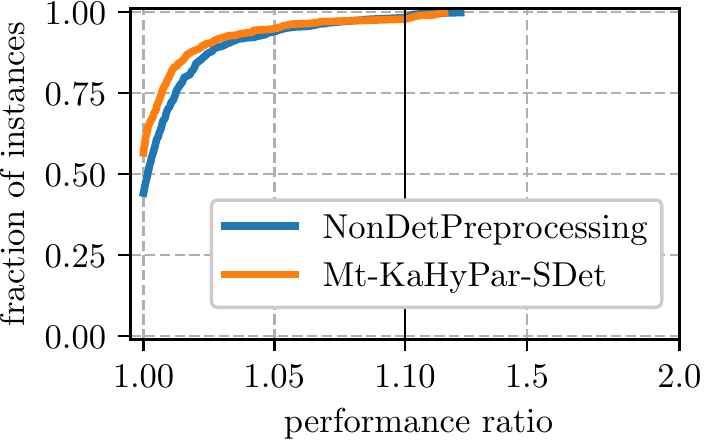}
	\end{minipage} %
	\begin{minipage}{.33\textwidth}
		\includegraphics[width=\textwidth]{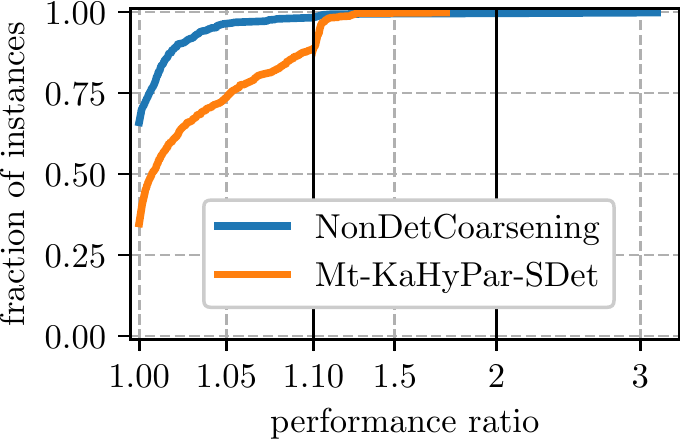}
	\end{minipage} %
	\begin{minipage}{.33\textwidth}
		\includegraphics[width=\textwidth]{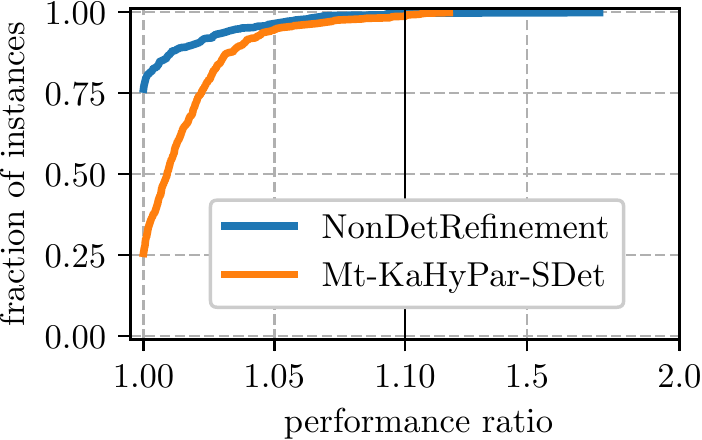}
	\end{minipage} %
	\caption{Performance profiles illustrating the quality impact of determinism in each component.}
	\label{fig:cost-of-determinism}
\end{figure*}

\begin{figure}
	\centering	\includegraphics[width=0.9\linewidth]{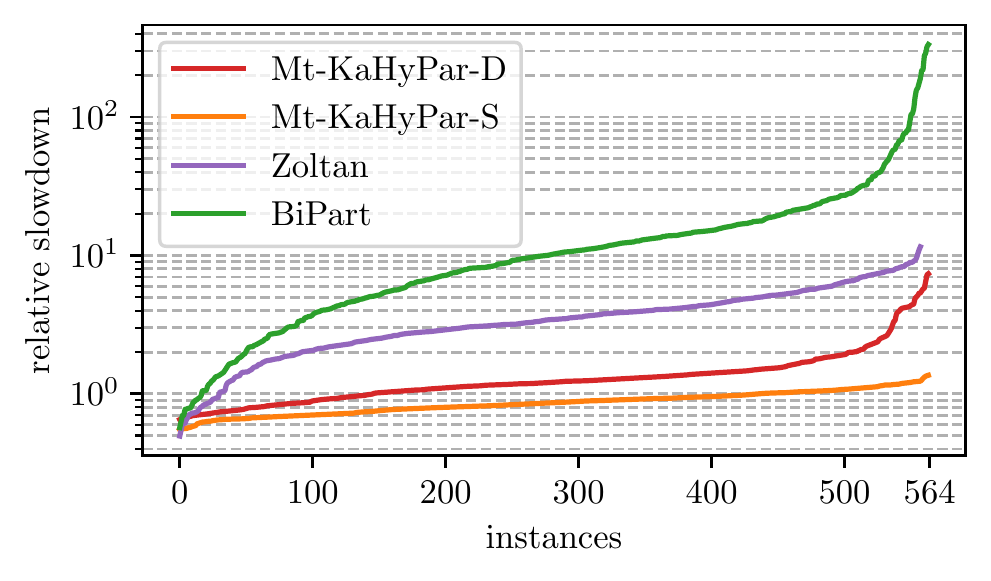}
	\caption{Slowdown of the other algorithms relative to \texttt{Mt-KaHyPar-SDet}, each using 64 threads. The instances on the x-axis are sorted independently for each algorithm.}\label{fig:relative-time:all-algos}
\end{figure}

\begin{figure}
	\includegraphics[width=\linewidth]{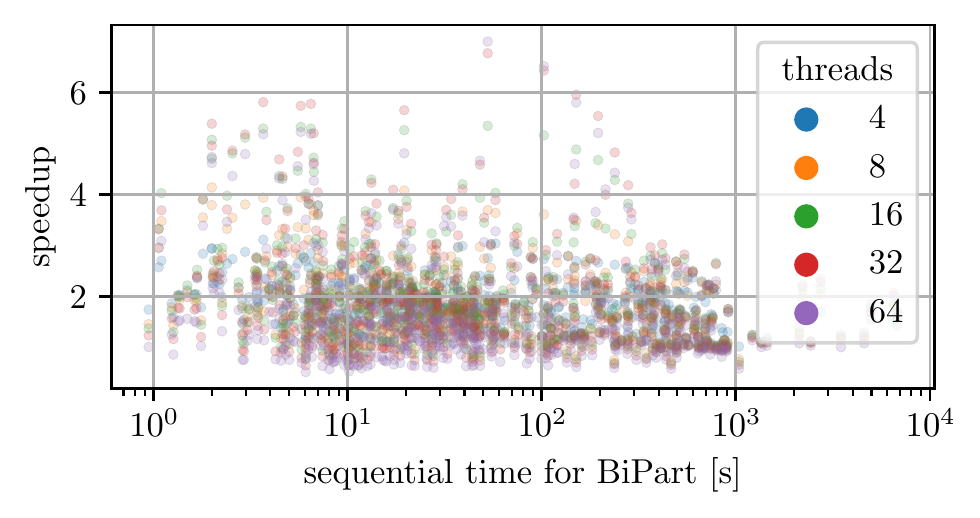}
	\caption{Speedups for \texttt{BiPart}.}\label{fig:speedups:bipart}
\end{figure}

The supposedly most important parameter is the number of sub-rounds used, as it offers a trade-off between scalability (synchronization after each sub-round) and solution quality (more up-to-date information).
In the following, we show that this is actually not a trade-off, as the number of sub-rounds either does not affect solution quality, or using fewer sub-rounds even leads to better quality.

\paragraph{Sub-rounds}

We made an initial guess of $5$ sub-rounds for refinement, and $16$ sub-rounds for coarsening and preprocessing, which we use as a baseline configuration when varying each parameter.
Figure~\ref{fig:subrounds:quality} shows the performance profiles.
The largest impact is on the coarsening phase, where $2$ sub-rounds performs the best.
Such a small value is surprising, yet one possible explanation is that high-degree vertices attract low-degree vertices too quickly if synchronization happens too frequently.
Using only $1$ sub-round is excluded here, since the clustering oscillates, which leads to coarsening converging long before the contraction limit is reached and thus initial partitioning takes very long.
Furthermore, using $2$ sub-rounds is about 12\% slower than using $3$ sub-rounds in the geometric mean, again due to the same effect.
Since it gives only slightly worse solution quality, we choose $3$ sub-rounds for coarsening in the main experiments.
For preprocessing, there is little impact on solution quality.
Here we stick with our original choice of $16$ sub-rounds since the floating-point-aggregation handling becomes substantially slower if more vertices are in a sub-round due to the sorting overhead.
For refinement, there is again little difference, where $1$ sub-round narrowly emerges as the best choice.
This is again surprising, as frequently synchronizing should allow for more informed move-decisions.
One cause we noticed is that with more sub-rounds the pair-wise swaps did not have sufficiently many moves to balance, as moves from earlier sub-rounds are not considered.
Using such moves as back-up could be included in future versions of the algorithm.

\paragraph{Impact of Preprocessing}

In Figure~\ref{fig:quality:no-preprocessing}, we show that the preprocessing phase is important for solution quality, which justifies the overhead for the floating-point volume updates.

\subsection{Scalability}

In Figure~\ref{fig:speedups:mt-kahypar-sdet} we show self-relative speedups of the overall algorithm and the separate components, plotted against the sequential running time on that particular instance.
In addition to the scatter plot, we show rolling geometric means with window size 50.
The overall geometric mean speedups of the full partitioning process are 3.91, 7.04, 12.79, 21.32, 28.73, 29.09 for 4, 8, 16, 32, 64, and 128 threads, respectively, and the maximum speedups are 4.9, 8.7, 15.8, 29.1, 48.9, and 72.6.
Since our algorithms are memory-bound workload types these are very good results.
On about 37\% of the runs with 4 threads, and 0.32\% of runs with 8 threads, we observe super-linear speedups which occur in all phases except initial partitioning.
We identified two reasons for this.
First, even sequential runs had running time fluctuations, and as super-linear speedups occur mostly for small sequential times, the speedups are more easily affected.
Second, while most of the work performed is deterministic, in all phases except initial partitioning we sort vectors that are filled in non-deterministic order.
Sorting algorithms have checks for pre-sorted sub-sequences to speed up execution.

Looking at speedups for the indidivual phases, we see that most phases exhibit very consistent speedups, even for small sequential running times.
Only initial partitioning exhibits sub-par speedups on larger instances, which is due to load imbalance from long running sequential FM refinement.

With 128 threads (only rolling geometric means shown for readability), the running times still improve, though not as drastically.
Only small instances show a slight slowdown, predominantly in initial partitioning.
Starting at $> 64$ threads, the second memory socket is used, so some slowdown is expected.
We use interleaved memory allocations to cope with NUMA effects as much as possible.

\subsection{Comparison with other Algorithms}

Figure~\ref{fig:quality:all-algos} shows performance profiles comparing our new algorithm \texttt{Mt-KaHyPar-SDet} with its non-deterministic variant \texttt{Mt-KaHyPar-S}, the stronger variant \texttt{Mt-KaHyPar-D} which uses parallel FM, the non-deterministic distributed algorithm \texttt{Zoltan}~\cite{ZOLTAN} as well as the deterministic \texttt{BiPart} algorithm~\cite{BIPART}.
In these experiments, each algorithm is run with 64 threads.
As expected, \texttt{Mt-KaHyPar-D} performs best, contributing the best solutions on about 75\% of the instances, followed by \texttt{Mt-KaHyPar-SDet} and \texttt{Mt-KaHyPar-S} which are similar, though \texttt{Mt-KaHyPar-S} is slightly better as it converges faster towards $1$.
\texttt{BiPart} is far off, contributing only 6 of the best solutions, and its quality is off by more than a factor of $2$ on more than 50\% of the instances; on some instances even by three orders of magnitude.
\text{Zoltan} is situated between \texttt{BiPart} and \texttt{Mt-KaHyPar-S}.
In a direct comparison, \texttt{Mt-KaHyPar-SDet} computes better partitions than \texttt{BiPart} on 551 of the 564 instances with a geometric mean performance ratio of 1.0032 compared to \texttt{BiPart}'s 2.3805.
In Figure~\ref{fig:relative-time:all-algos}, we report relative slowdowns, i.e., the running time of the other algorithm divided by running time of the baseline \texttt{Mt-KaHyPar-SDet}.
\texttt{Mt-KaHyPar-S} is faster on all but 158 instances and never by a factor of more than $2$.
\texttt{BiPart} is between one and two \emph{orders of magnitude} slower than the two speed variants of \texttt{Mt-KaHyPar}.
The reason for this is shown in Figure~\ref{fig:speedups:bipart} which shows self-relative speedups of \texttt{BiPart} for increasing number of threads.
Most speedups are below 2 and the largest speedup is about 7.

\subsection{The Cost of Determinism}

In this section, we investigate in which phase the solution quality of \texttt{Mt-KaHyPar-SDet} gets lost, by swapping out one component for its non-deterministic counterpart, in each of the plots in Figure~\ref{fig:cost-of-determinism}.
Interestingly, the biggest quality loss comes from coarsening, whereas deterministic preprocessing even improves quality.
The loss in refinement is expected due to the lack of up-to-date gains and the inability to leverage zero gain moves for rebalancing and diversification~\cite{mt-kahypar-d}.

For coarsening, the results are unexpected, particularly because similar local moving algorithms~\cite{SLM} are not as affected by out-of-date gains.
One reason for this is that multiple global rounds are performed, where vertices can back out of their first cluster assignment.
We only use one round and even prematurely terminate the round to avoid coarsening too aggressively.
Performing a second round, where only already clustered vertices may reassess their assignment, may be beneficial and we leave this for future research.
Additionally, we unsuccessfully experimented with several features of the non-deterministic coarsening such as adapting hyperedge sizes to the current clustering in the rating function and \emph{stable leader chasing}, where oscillations (vertices joining each other) and cyclic joins are resolved by merging all involved vertices.

\section{Conclusion and Future Work}\label{sec:conclusion}

We presented the first scalable, deterministic parallel hypergraph partitioning algorithm.
Our experiments show that determinism does incur sacrifices regarding both solution quality and running time compared to the previous non-deterministic version, but these are small enough to justify if determinism is desirable.
Future work includes incorporating determinism into additional refinement algorithms, improving performance on multi-socket machines, and implementing these techniques for distributed memory.

For example for flow-based refinement~\cite{REBAHFC,KAHYPAR-HFC} this is well within reach, as scheduling on block-pairs can synchronize after each block was involved in a refinement step, and the flow algorithms need not be deterministic since the used cuts are unique.
Parallel localized FM seems like a much more difficult target, though a promising approach may be to stick with approving expansion steps at synchronization points.
Additionally, handling extremely large $k$ and speeding up initial partitioning is possible by employing the deep multilevel approach~\cite{deep-mgp} instead of recursive bipartitioning during initial partitioning.

\begin{acks}
  This work was supported by the Deutsche Forschungsgemeinschaft (DFG, German Research Foundation) under grants WA~654/19-2 and WA~654/22-2.
\end{acks}

\bibliographystyle{ACM-Reference-Format}
\bibliography{partitioning_references}

\appendix

\end{document}